\documentclass[twocolumn,showpacs,preprintnumbers,amsmath,amssymb]{revtex4}
\usepackage{amssymb}
\usepackage{color}

\usepackage{graphicx}
\usepackage{dcolumn}
\usepackage{bm}

\begin{document}

\title{A Momentum Filter for Atomic Gas}

\author{Wei Xiong}
\author{Xiaoji Zhou} \thanks{Electronic address: xjzhou@pku.edu.cn}
\author{Xuguang Yue}
\author{Yueyang Zhai}
\author{Xuzong Chen} 
\affiliation{School of Electronics Engineering $\&$ Computer
Science, Peking University, Beijing 100871, China}
\date{\today}

\begin{abstract}
We propose and demonstrate a momentum filter for atomic gas based
on a designed Talbot-Lau interferometer. It consists of two identical
optical standing-wave pulses separated by a delay equal to odd multiples of the
half Talbot time. The one-dimensional momentum width along the long direction
of a cigar-shaped condensate is rapidly and greatly purified to a minimum, which
corresponds to the ground state energy of the confining trap in our experiment.
We find good agreement between theoretical analysis and experimental results.
The filter is also effective for non-condensed cold atoms and could be applied
widely.
\end{abstract}

.

\maketitle
\section{Introduction}
Atomic sources with long coherence lengths, corresponding to a narrow momentum width,
can be used to observe new physical phenomena and contribute to the improved precision
measurement spectroscopy. Atomic clocks~\cite{Mark,Wynands} are for example greatly improved with
narrower momentum width and so are the atomic interferometers~\cite{Alexander,Fixler}. Large correlation lengths have been achieved with Bose-Einstein condensate~\cite{Anderson,Davis2}. Many techniques have been used to actually reduce the momentum width of atomic gases. Some of them, such as the
velocity-selective coherent population trapping (VSCPT)~\cite{Aspect} and Raman filter~\cite{Kasevich}, behave as momentum filters to select atoms with specific momenta and to discard the others. Other techniques benefit from the thought of filter to achieve lower temperature, such as evaporative cooling~\cite{Ketterle,Davis} and Raman cooling~\cite{Kasevich2}.

In this paper, we report a momentum filter scheme by precisely
discriminating the different momenta-related phase evolutions during
a matter wave Talbot-Lau interference sequence. The Talbot effect
was observed as a near-field diffraction with optical
waves~\cite{William} and later observed with
atoms~\cite{Michael,Clauser,Nowak,Clauser2,Carnal,Deng}. In those
works on the matter wave Talbot effect, the initial matter waves
were approximated as a mono-energetic plane wave. In our work, the
momentum distribution of a practical matter wave is considered. The
effect of the momentum filter appears when an atomic gas is
diffracted by a designed temporal Talbot-Lau interferometer with
specific standing wave pulses and time intervals between the pulses.
As a result, the interferometer generates different interference
patterns for different initial momenta and behaves as a filter. When
the filter is applied on condensates, we choose the intervals to be
short enough (less than $11T_T/2$), so that the momentum filtering
affects the scattering process more than the interaction energy induced
by the interaction between atoms. The redistributing and purifying
effects of the filter are demonstrated in the experiments and also
analyzed theoretically. With the filter, we are able to rapidly purify one dimensional momentum width of the atomic gas to a minimum, which is the momentum width of atoms in the ground state of the confining trap. Additionally, we discuss the influence of atomic
interaction in the experiments above. The momentum filter is
effective for both non-condensed cold atoms and condensate as shown in the
experiment. We also discuss some possible applications of the
filter.

This paper is organized as follows. In Sec. II, a physical model to
interpret the momentum filter is presented. We derive a concise
expression with Raman-Nath approximation~\cite{Edwards} to reveal
the physics of the filter, and then treat the standing wave as a one
dimensional optical lattice for more precise description. We also
discuss the effect of interaction between atoms for filters on
condensates. Section III and IV present the effect of momentum
redistribution and purification by the filter respectively. In Sec.
V, the flexibility of the filter is demonstrated. Section VI
contains discussion and conclusions.

\section{Physical model}
\begin{figure}
\centering
\includegraphics[width=8.5cm]{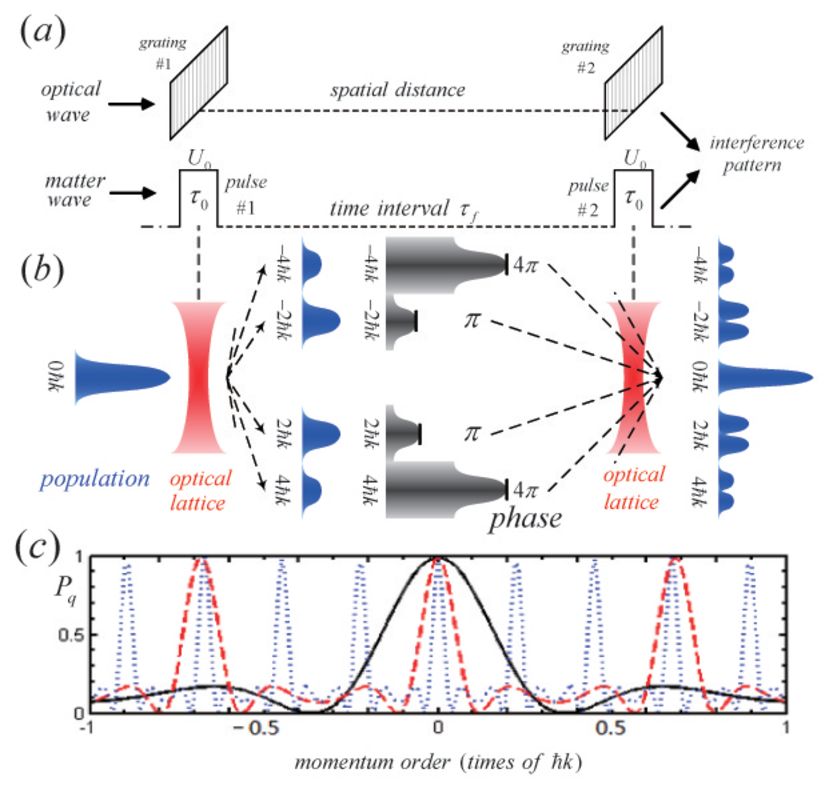}
\caption{(color online) The temporal Talbot-Lau effect on matter
wave as a momentum filter. (a) The similarity between the optical
and the matter wave Talbot-Lau effect. The input wave source is an
optical (matter) wave. The two standing wave pulses behave as the
two gratings. The phase of the light wave evolutes along the space
distance similarly as the matter wave during the time interval. (b)
The scheme of the momentum filter. An atomic cloud with a momentum
distribution around zero is scattered to be around momenta $\pm 2n
\hbar k$ ($n=0,1,2,...$) by the first pulse. During a time delay
$T_T/2$, the atoms with momenta centered around $\pm 2n \hbar k$
acquire a phase shift around ${n}^{2}\pi$. After the second pulse,
the atoms with the initial momenta closer to zero will be more
probably scattered back into the initial state. (c) The relation
between the probability $P_q$ and its initial momentum $q$. The
solid, dashed and dotted line correspond to the interval of $T_T/2$,
$3T_T/2$ and $9T_T/2$.} \label{phy}
\end{figure}

Similarly to the optical Talbot-Lau effect, the temporal Talbot-Lau
effect on matter wave behaves as shown in Fig.~\ref{phy}(a). Two
standing wave (optical lattice) pulses with the same duration
$\tau_0$ and induced optical potential $U_0$, behave as two
gratings. The matter wave accumulates a phase shift during the time
interval $\tau_f$ as what happens to a light wave along its optical
path. The two effects both lead to the matter (optical) wave
interference patterns.

Fig.~\ref{phy}(b) shows the mechanism of a band-pass momentum filter
centered at zero momentum. The time interval $\tau_f$ is
designed as odd multiples of the half Talbot time $T_T/2$, where the
Talbot time $T_T$~\cite{Deng} is the minimum time for the matter
wave front's reconstruction after the first pulse. An atomic gas
with its momenta distributed around zero is diffracted into the
components with momenta around $\pm 2n \hbar k$ ($n=0,1,2,...$, the
Plank constant $\hbar$, the wave vector of the light for the
standing wave $k$) by the first pulse. With the interval $T_T/2$,
the atoms with momenta $\pm 2n \hbar k$, which originate from the
atoms with initial zero momentum, obtain the accurate phase
evolution $n^2\pi$ (which can be simplified as $n\pi$ because $n^2$
and $n$ have the same parity), and the second pulse entirely
diffracts them back to their initial states. For the atoms with
non-zero initial momenta, the larger the initial momentum is, the
larger the phase deviation from $n^2\pi$ is and the more imperfect
the recurrence will be, which results in a momentum filtering.

For a brief description of the filter, we consider the standing wave
pulses to be short enough for the Raman-Nath approximation and
obtain the probability of an atom with initial momentum $q$
returning to the initial state as:
\begin{equation}\label{pq}
{P_q} = \sum\limits_{n =  - \infty }^{ + \infty } {{J_{ - n}}}
({\Omega}\tau_0 ){J_n}({\Omega}\tau_0 ){e^{ - i{E_{n,q}}{\tau
_f}/\hbar }},
\end{equation}
with the Bessel function of the first kind $J_n$, the two-photon
Rabi frequency $\Omega$~\cite{Linskens}, the kinetic energy
${E_{n,q}} = {(2n\hbar k + q)^2}/2M$ and the atomic mass $M$. While
the interval is designed as $\tau_f = (2N+1)T_T/2, (N=0,1,2...)$,
the atoms with initial zero momentum accumulate the phase as
$E_{n,0}{\tau _f}/\hbar  = {n^2}(2N + 1)\pi $, and the probability
reaches the maximum as ${P_{q = 0}} = 1$.

Afterwards, we apply the Bloch theorem for a more precise description
of the scattering process as described in ~\cite{xiong}, where the
standing wave is considered as an optical lattice and re-calculate
the probability $P_q$ as shown in Fig.~\ref{phy}(c). It can be seen
from the figure that the momentum filter actually consists of a
series of filters with different centers. The filter can be
described as long as the width and position of each filter are
definite. The $1/e$ width $2q_0$ of a single filter can be evaluated
based on ${P_{q_0}} = 1/{e}$. Since the probability is principally
influenced by the phase evolution during the interval, we introduce
a phase shift index $\alpha$, which satisfies $\alpha \pi  =
({E_{1,{q_0}}} - {E_{1,0}}){\tau _f}/\hbar  \approx 2k{q_0}{\tau
_f}/M$, while ${q_0} \ll 2\hbar k$ is easily satisfied. The index
$\alpha$ indicates the phase deviation between the atoms with the
momentum $2\hbar k$ and $2\hbar k+q_0$ during the interval. It
reveals the relation between the interval and the width of a single
filter as:
\begin{equation}\label{width}
  2{q_0} \approx M\alpha \pi /k{\tau _f}
\end{equation}
Since $\tau_f$ is equal to $(2N+1)T_T/2$, the width is also of
discrete values. As shown in Fig.~\ref{phy}(c), when the interval is
increased, the width of the filter decreases and so does the
distance $\delta q$ between adjacent centers. This distance means
that the atoms with an initial momentum difference $\delta q$ will
obtain the same interference pattern after the filter, since the
phase deviation $({E_{n,q + \delta q}} - {E_{n,q}}){\tau _f}/\hbar $
is integral multiple of $2\pi$. According to our analysis, $\delta
q$ satisfies $\delta q = 2\hbar k/(2N+1)$, which can also be drawn
from the figure.

We analyze the momentum filter with single-atom model above, which
works for non-condensed cold atoms. When the atomic source is a condensate,
interaction between atoms needs to be considered. When condensates
are scattered by one dimensional filter sequence, the interaction
between atoms takes effect. The widths of light pulses (several $\mu
s$ generally) are much shorter than the characteristic time of
interaction energy (magnitude as $ms$), so interaction during light
pulses can be neglected. Atomic interaction can not be ignored so
easily during intervals between light pulses, because intervals are
possibly comparable with the characteristic time of the interaction
shift. For a condensate in a harmonic trap exposed to a
one-dimensional momentum filter, its interaction energy can be
averaged in the plane perpendicular to the lattice direction and get
a distribution along the lattice direction as $E_{I}={E_{0}}(1 -
{z^2}/{R_Z}^2),$ with the interaction energy maximum $E_{0}$
in lattice direction $Z$, the coordinate of lattice direction $z$
and the Thomas-Fermi radius $R_Z$ of the condensate along this
direction.

According to the Gross-Pitaevskii equation, the phase of $2\hbar k$
component after the first lattice pulse evolves as ${e^{ - i({E_{k}}
+ {V} + {E_{I}}){\tau _f}/\hbar }}$, with the kinetic energy
${E_{k}}$ along Z direction and magnetic trap $V$. The frequency of
the magnetic trap along $Z$ direction is only $20Hz$, so it can be
neglected, compared with the other two components. The kinetic energy
is related to the Talbot time as $E_{k}/(2\pi \hbar) = 1/T_T$. For the
initial momentum $q$ corresponding to a momentum width introduced by
momentum filter, the rate of its induced phase shift is ${(2\hbar k
+ q)^2} - {(2\hbar k)^2}/2Mh \approx 2\hbar k \times q/Mh$.

On the other hand, for interaction between atoms along the Z direction,
the average of interaction energy is $80\%$ of the maximum. The rate
of the phase shift induced by interaction can be approximated as
$E_{0}/5h$. If the rate of the phase shift induced by the initial
momentum is larger, i. e. $2\hbar k \times q/M
> {{\rm{E}}_{0}}/5$, momentum filter plays a main role in the
scattering process. Otherwise, interaction energy plays a major role.

In our experiment, the Talbot time is $T_T=76\mu s$, and the initial
momentum $q$ leads to a rate of phase shift as $q/\hbar k \times
13.16kHz$ . The interaction energy is about $1.5kHz$ at the center of
the atomic cloud in our experiment and its distribution along $Z$
direction after average is $1kHz \times (1 - {z^2}/{R_Z}^2)$ . The
rate of phase shift induced by atomic interaction is about $200Hz$
in our experiment. While interval increases, the corresponding momentum width decreases, and the phase evolution rate relevant to the momentum width also decreases. On the other hand, the phase evolution rate induced by interaction is independent of the interval, so these two rates can be equal with a certain interval. When the rates introduced by initial momentum and
interaction are equal, the interval is $6T_T$ and it means that the
effect of momentum filter plays bigger role than the interaction
when the interval is shorter than that, and things are opposite when
the interval is longer. The interaction induced phase shift is
position dependent and the interaction energy will accelerate the
phase evolution of atoms around the center of atomic cloud. The
filtered fraction results of both momentum filter and interaction
energy effects. As a result, we demonstrate the momentum filter
mainly within the regime, where intervals are less than $6T_T$.

\section{Momentum redistribution}
We performed the experiments on a $^{87}$Rb Bose-Einstein condensate
(BEC) system~\cite{zhou,lu}. After pre-cooling, a cigar shaped
$^{87}$Rb condensate of $1\times10^5$ atoms in $5{}^2{S_{1/2}}\left|
{F = 2,{M_F} = 2} \right\rangle$ state was achieved by radio
frequency (RF) cooling in the a harmonic magnetic trap (MT), with
$20$~Hz axial frequency and $220$~Hz radial frequency. The initial
atomic gas is prepared as a condensate without observed non-condensed cold atoms. A
one dimensional designed far-red-detuned (the wavelength is $852nm$)
standing wave pulse sequence was applied onto the condensate along
the axial direction. The lattice depth is $80E_R$, which is
calibrated by Kapitza-Dirac scattering experimentally. In the
experiments, we choose the width of a single pulse to be $3\mu s$,
so that the first pulse is able to couple most of the atoms (about
$60\%$) to the momenta $\pm 2\hbar k$, some atoms are populated
around $\pm 4\hbar k$ (about $36\%$), others are around $0\hbar k$
and other momenta. The second pulse is the same as the first one. In
our experiment, the Talbot time is $76\mu s$ and the index $\alpha$
is about $0.20$. According to our calculation, the influences from
momentum filter and atomic interaction are equal when the interval
is $6T_T$, so we choose the interval from $T_T/2$ to $11T_T/2$ in
experiments, where momentum filter plays bigger role.

\begin{figure}
\includegraphics[width=8.5cm]{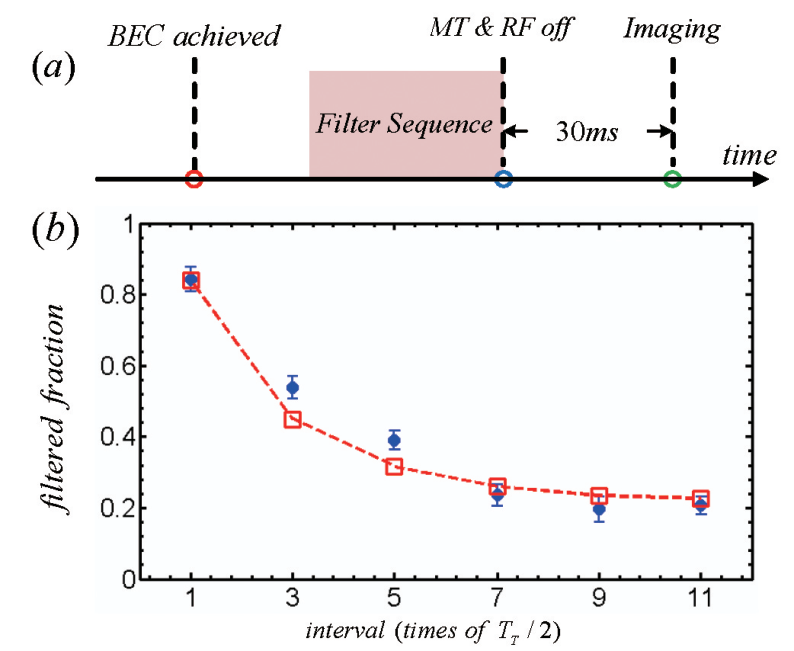}
\caption{(color online) Redistributing effect of the momentum
filter. (a) Sequence of the experiment. (b) The ratio of the
filtered atoms ($N_{0}$) over the total number of the atoms ($N_A$)
versus the interval $\tau_f$. The round solid dots are the
experimental results and the hollow square dots are the theoretical
analysis, considering that the effect of atomic gases' practical
temperature and the momentum expansion induced by $s$ wave
scattering are equivalent to an initial momentum width $0.1\hbar k$.
} \label{filter}
\end{figure}

We demonstrate the redistributing effect of our momentum filter by turning off the RF field and the magnetic trap simultaneously, after sending the pulse sequence. The time-of-flight (TOF) signal of the atoms is obtained after $30ms$ free fall (see Fig.~\ref{filter}(a)). As the interval increases, the bandwidth of the filter decreases and the relative population of the atoms returning to their initial states also decreases as shown in Fig.~\ref{filter}(b). For the non-condensed-cold-atom background of the cloud, its momentum distribution is classically Gaussian. However, for the condensate, when its momentum can not be neglected, Thomas-Fermi approximation is no longer valid and it is reasonable that its momentum distribution is analogous to Gaussian profile. Therefore, we describe condensate's momentum distribution approximatively as $c{e^{ -  {q^2}/{\Delta ^2}}}$ (the normalization coefficient $c$, the momentum $q$ and the $1/e$ width $2\Delta$). In this way, we calculate the relative population (square hollow dots in the figure) of the filtered atoms, compare the calculation with the experiment results (round solid dots) and find that the calculation generally pictures the trend of experimental results. The decrease of filter bandwidth, derived from Eq.~(\ref{width}) as $\delta=4\alpha \hbar k/(2N + 1)(2N + 3) \approx \alpha \hbar k/{N^2}$, shrinks rapidly as a function of $N$. As a result, the number of the filtered atoms changes more and more slowly while the interval $\tau_f$ increases by units of $T_T$ in Fig.~\ref{filter}(b).

For the theoretical curve (see square hollow dots in
Fig.~\ref{filter}(b)), atomic gases to be filtered are not pure
condensates with zero temperature and $s$ wave scattering
~\cite{band} also introduces momentum dispersion during the recoil
after the first pulse. We approximate these two effects as an
average initial momentum width $0.1\hbar k$~\cite{xiong}. When the
intervals are short, the simulated filtered fractions are slightly
higher than experimental ones, because the effective initial
momentum widths are lower than $0.1\hbar k$. While the intervals get
longer, the effective initial momentum widths are higher, so the
filtered fractions are lower.

There are still some deviations between the calculation and the
experiment, although the experimental results are normalized to
minimize the uncertainty of the initial atom number. One of the
possible reasons may be the assumed purity of the condensate in the
theory. A practical condensate is always surrounded by a cloud of
non-condensed cold atoms with much wider momentum width. The non-condensed cold atoms contribute
little to the filtered component, but much more to the total atom
number. As a result, the fluctuation of the non-condensed cold atom number leads to
the deviation in our experiment with respect to theory.


\begin{figure}
\includegraphics[width=8.5cm]{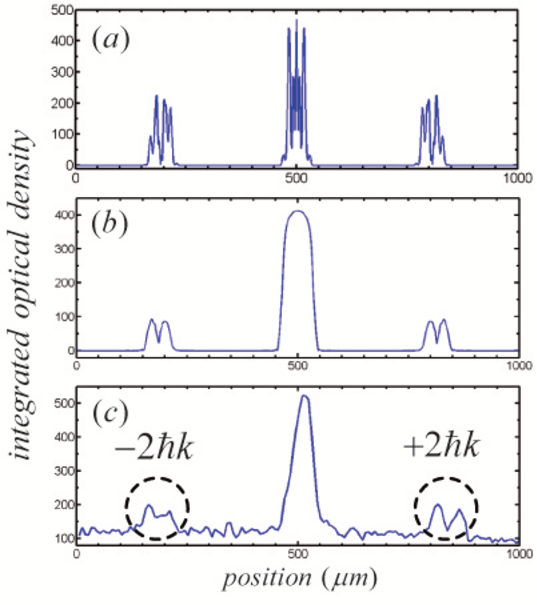}
\caption{(color online) Redistribution introduced by both momentum
filter and atomic interaction. (a) A numerical simulation of TOF signal considering momentum filter without interaction. (b) A numerical simulation of TOF signal including momentum filter and atomic interaction. (c)  A TOF signal obtained in our experiment. Two holes around $\pm 2\hbar k$ components can be observed.} \label{interaction}
\end{figure}

In the experiment about momentum redistribution by a pulse sequence, when the interval $\tau_f$ is $17T_T/2$, we observed two holes at the position of the momenta $\pm 2\hbar k$ in TOF signal as shown in Fig.~\ref{interaction}(c). With this interval, the rate of phase shift induced by momentum corresponding to the width of the filter is $141Hz$ and that induced by interaction is still $200Hz$, so that the two holes originate from both position selection by atomic interaction and momentum filtering. We numerically simulated the process with the only difference that the initial condensate is with (see Fig.~\ref{interaction}(b)) or without (see Fig.~\ref{interaction}(a)) atomic interaction. The two simulations clearly show the effect introduced by atomic interaction.

As shown in Fig.~\ref{interaction}(a), for condensate without atomic interaction, many peaks appear in the component around $0\hbar k$, and each peak corresponds to a certain center of the filter, because the absence of atomic interaction leads to little expansion in TOF signal and none of other redistributing effects. Consequently, only the momenta parts produced by the filter can be seen and the valleys in the  $\pm 2\hbar k$ components also illustrate that. In Fig.~\ref{interaction}(b), for condensate with atomic interaction, only one peak emerges around $0\hbar k$, and it is because that interaction leads to expansion and blur the peaks in TOF signal. However, two valleys arise around $\pm 2\hbar k$ separately. They are wider than the ones in condensate without interaction and obviously come from phase gradient introduced by atomic interaction, except for the filter. We numerically analyzed the process and found that the result is generally consistent with the experimental result. The holes in Fig.~\ref{interaction}(c) are even larger, and the reasons may be dispersion induced by s-wave scattering, deviation from imaging system and so forth.


\section{Momentum Purification}
\begin{figure}
  \centering
  \includegraphics[width=8.5cm]{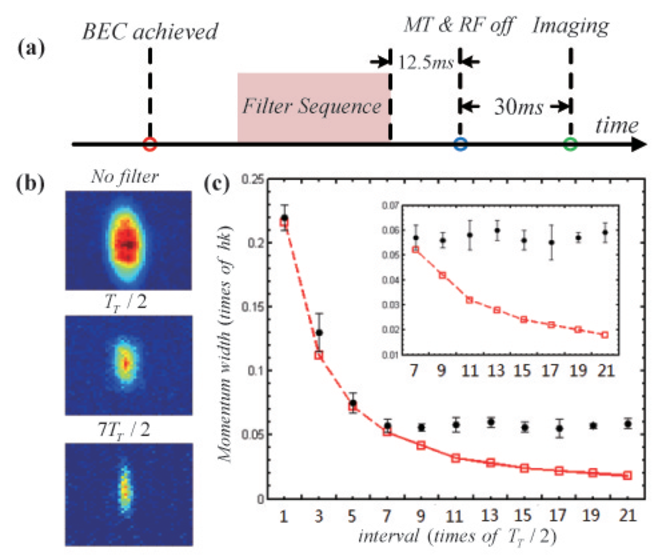}
\caption{(color online) Purification effect of the momentum filter.
(a) Time sequence for momentum narrowing. (b) TOF signals
demonstrating one dimensional momentum narrowing. The figures from
top to bottom show a condensate without filtering, a condensate
modified by a filter with interval $T_T/2$ and a condensate
purified by a filter with interval $7T_T/2$ respectively. (c)
Momentum width versus the interval of the filter. The hollow square
dots are theoretical results and the solid round dots are
experimental results.} \label{compress}
\end{figure}

To observe the momentum purifying effect of the filter, we manage
to keep the filtered component in the magnetic trap and discard the
others. As described in Fig.~\ref{compress}(a), the magnetic trap
and the RF field are maintained for $12.5$~ms (a quarter of the
period of the magnetic trap along the axial direction) after the
filter sequence. The atoms with higher momentum will move away to
higher potential positions of the magnetic trap and be driven out by
the RF field. On the other hand, the remaining atomic gas kept in
the trap evolves during this time. The $1/e$ width $Z(0)$ of the
atomic gas at the moment of being released from the magnetic trap
can be calculated based on the wave function of the condensate after
the filter. The evolution of the atomic gas can be figured out
according to the time-dependent Gross-Pitaevskii equation. We are
able to evaluate the momentum width after the free falling time $t$
from the TOF signal by working out the ballistic expansion equation
${Z^2}(t) = {Z^2}(0) + {({2q_0}/M)^2}{t^2}$, with the size $Z(t)$ of
the cloud along the lattice direction.

The TOF signal in Fig.~\ref{compress}(b) demonstrates the one
dimensional momentum purification with increased interval. The
reduction of the atomic cloud size along the lattice direction from
the TOF signals pictures clearly the effect of the momentum
purification. The figures from top to bottom show a condensate
without filtering, a condensate modified by a filter with interval
$T_T/2$, a condensate purified by a filter with interval $7T_T/2$
respectively. The atomic size along the lattice direction in the
figure at the bottom is the narrowest one we can get in the
experiment.

The atomic cloud with the narrowest size we obtained in experiment was not in the Q2
Thomas¨CFermi regime. The atomic cloud contained about $2\times 10^{4}$ atoms. If it reached thermal
equilibrium and were in the Thomas-Fermi regime, its initial size in
the trap along $Z$ direction would be $2{R_Z} = {(2{\mu _N}/m{\omega
_Z}^2)^{1/2}} \approx 52\mu m$ ($\mu_N$ is the chemical potential
related to s-wave scattering length $a$, atomic number $N_0$ and
geometrical frequency of the magnetic trap $\omega$ as ${\mu _N} =
{(15{N_0}{\hbar ^2}a/(4\sqrt 2 M))^{2/5}}{(M{\omega ^2})^{3/5}}$ ).
After $30ms$ expansion, the size of atomic cloud would reach $60\mu
m$, which is much larger than $28\mu m$ as observed in experiment.
As a result, the size of atomic cloud's shrinking mainly depends on
momentum purifying, but not the decrease of atomic number.

The achieved one dimensional momentum width by the filters with
increased interval is shown in Fig.~\ref{compress}(c). The momentum
width predicted by the calculation (square hollow dots) goes along
with the experimental results (round solid dots) well until the
interval of the filter exceed $7T_T/2$, which corresponds to the
expected width $0.052\hbar k$. When the atomic gas is abruptly released from the magnetic trap, its potential energy vanishes but the kinetic half remains, so the average kinetic energy of the atomic gas in free space can not get lower than that of the ground state of the confining trap. The
frequency $20Hz$ of the axial magnetic trap corresponds to the
momentum width $0.056\hbar k$ as $(0.028\hbar k)^2/M = h \times
20Hz/4$ and this is the reason why the momentum width of the atomic
gas keeps this value, although a filter with longer interval is in
principle able to further purify the momentum width.

We can still observe narrow size of atomic gas, even with very long
interval (up to $21T_T/2$). The reason is as follows. Although
interaction is present during the process, it only leads to an
imperfection of the momentum filter as it decreases the probability
of atoms with initial momentum around zero to return to their
original states. The momentum filter still purifies the momentum
width as anticipated. The interaction energy is position dependent,
not momentum dependent. It affects the population of the
momentum-filtered component, but hardly the momentum width. On the
other hand, the ground state energy of our magnetic trap is a
stronger constraint to the momentum width of the released atomic
gas. We observe the narrowest kinetic energy dispersion of the
released atomic gas corresponds to the ground state energy of the
trap, achieved for a filter using a $7T_T/2$ interval. When
intervals are longer than that, neither the effect of the momentum
filter nor that of the interaction energy can be observed on the
momentum widths. As a result, we can still observe atomic clouds
shrinking to the narrowest size with very long intervals in our
experiments.

\section{Flexibility of the filter}
We show in this section that the filter can be applied flexibly, as
it can be used for both non-condensed cold atoms and condensate. Several filters
can be combined to a new one and the center of a filter may also be
adjusted.

\begin{figure}
  \centering
  \includegraphics[width=8.5cm]{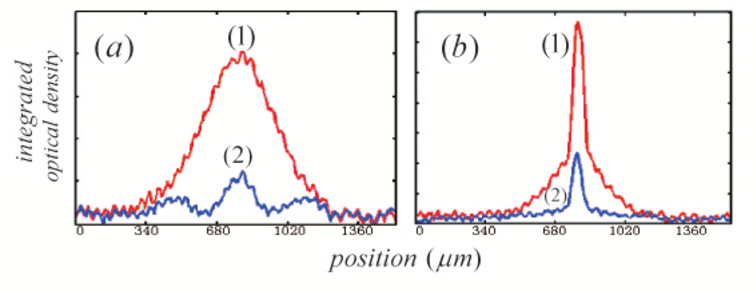}
\caption{(color online) TOF signals showing the filtering on two
kinds of atomic gases. Curve (1) is the initial atomic gas and curve
(2) is the atomic gas after the filter. (a) A cloud of non-condensed cold atoms with temperature $513nK$
modified by a filter with the interval $T_T/2$. (b) A mixture of condensate and non-condensed cold atoms with temperature $346nK$ modified by a filter with the interval
$5T_T/2$.} \label{filter2}
\end{figure}

The effectiveness of the filter for different kinds of atomic gases
is demonstrated in Fig.~\ref{filter2}. The effect of a filter for
non-condensed cold atoms is shown in Fig.~\ref{filter2}(a) with the interval
$T_T/2$ and that of a filter with the interval $5T_T/2$ on a mixture
of condensate and non-condensed cold atoms is shown in Fig.~\ref{filter2}(b). The
re-shaping of the momentum distribution can be obviously observed
from the TOF signal. Actually, there is no restriction with respect
to the choice of the initial atomic cloud for the filter, but the
parameters of the filter still need to be optimized for the
observation of the filter effect.

\begin{figure}
  \centering
  \includegraphics[width=8.5cm]{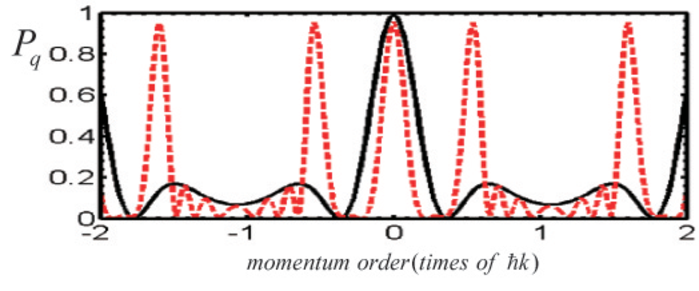}
\caption{(color online) Comparison between single filter and two
combined filters. The solid curve is the probability of recurrence
for the single filter with interval $T_T/2$, the dashed one is that
for a combination of two same filters with interval $T_T/2$ linked
by a $T_T/2$ time delay.} \label{combine}
\end{figure}

Combination of single filters is a possible choice for achieving
narrower width. The probabilities of recurrence after a single
filter with the interval $T_T/2$ and two combined same filters are
compared in Fig.~\ref{combine}. The probability of the combined one
is not simply the production of two single filters, because for the
second filter sequence, the filtered component is the superposition
of the states with a series of different phases instead of a pure
state with the same phase as the initial state; additionally, the
components with the momenta $\pm 2n\hbar k$ still remain in the
atomic gas and play a role in the interference of the second filter
sequence. However, the combination still behaves as a filter with a
narrower momentum width.

The center of the momentum filter is not confined to be only zero,
such as some value $q_c$, we can design the filter with time
constant $\tau_f$ as ${E_{n,{q_c}}}{\tau _f}/\hbar = (2N + 1)\pi $
($N=0,1,2,...$) with similar bandwidth. For instance, if the intervals are $2T_t/2, 4T_t/2, ..$, the minimum centers of the corresponding filters will be $\hbar k/2, \hbar k/4, ..$.

\section{Discussion and conclusions}
This momentum filter actually utilizes the interference among all the momentum states generated by the first lattice pulse to form a momentum filter. Although the $\left| { \pm 2\hbar k} \right\rangle$ and $\left| { \pm 4\hbar k} \right\rangle$ states overwhelm the others after the first pulse, the filter will be less functional if the other momentum states are neglected, let alone the $\left| { \pm 4\hbar k} \right\rangle $ states also being canceled. The consequence of ignoring higher orders' momentum states is that the filter will miss some zero momentum atoms and collect more non-zero components and the momentum distribution will be broadened.

The momentum width achieved by the filter may be limited by the
following factors. The first one is the initial momentum width of
the atomic gas. The initial momentum distribution can cover several
centers of the filter. Although the increased interval can reduce
the bandwidth, the non-zero centers will weaken the purifying of
the momentum. The second one is the relative line width $\gamma$ of
the laser of filter, since the interval can not increase for failure
of discriminating the phase shift due to the line width, which means
$({(2\hbar k(1 + \gamma ))^2} - {(2\hbar k)^2}){\tau _f}/2M\hbar \le
\alpha \pi $. The third one is the de-coherence time. The filter is
a process of interference, so there will be no filter if the quantum
states de-coherent during the interval of the filter.

In our experiment, the narrowest momentum width achieved with this
filter is limited by the confining trap. If the confinement of the
trap is weaker, the momentum filter will be able to generate
narrower momentum widths. The experiments can be extended to three
dimensions, since three dimensional optical lattice is widely used
at present. The three dimensional filter could be a tool for rapidly
cooling atomic gases.

This filter is not a repetitive demonstration of Talbot-Lau interferometer, since the matter wave we dealt with is distributed in momentum space, not a simple plain wave. As a result, the momentum selecting effect comes out from the interferometer. Briefly compared with other momentum filters, our filter is simple and robust.

Firstly, requirements on lasers are different in our filter and other velocity selection
methods. Our filter consists of optical lattice pulse sequences, which are easily designed and
controlled. The lattice laser is so far detuned from atomic resonance that the filter can work
well without laser frequency lock. This means a good many kinds of atoms and molecules
can be momentum purified by lasers with a very wide range of frequencies, through our
method. By comparison, optical setups a for Raman filter and VSCPT are more complicated.
For Raman filter, in order to preserve the atoms with specific momenta and to blow off the others, at least two different frequency laser beams are needed. Otherwise, the selected atoms
have to be charged a momentum $2\hbar k$. Furthermore, laser frequencies have to be locked to make
sure that Rabi frequency matches the pulse duration.

Secondly, our filter preserves both the internal and external states of selected atoms. It
hardly changes the momenta or the internal states of the selected atoms, and simultaneously
charges other atoms with momenta $\pm 2\hbar k$, $\pm 4\hbar k$ and so on. Consequently, our filter can work solo and cooperate with other techniques smoothly, such as evaporative cooling. For Raman
filter, if the internal states of filtered atoms need to be preserved, they have to be charged with
a momentum $2\hbar k$; if the momenta of the selected atoms are preserved, they have to be pumped
into another internal state. Otherwise, the atoms both preserving momenta and internal states
are difficult to distinguish from others.

Thirdly, our filter is robust for atoms being ¡®in the dark¡¯ during most of the filter process.
When lattice is on, lasers in our filter can work well with very large detuning (taking $^{87}Rb$ for
example, its $D1$ line resonates at $795nm$, our lattice light works at $852nm$) and very short pulses
(in our experiment, pulse duration can be as short as $1\mu s$ while lattice depth reaches $120E_R$);
during the rest of the filter process (in our experiment, the interval between two lattice pulses is
about $300\mu s$ to achieve the momentum width $0.056\hbar k$), atoms stay ¡®in the dark¡¯. As a result, the filter can hardly be disturbed by most of the perturbations and drifts on lattice¡¯s frequency, phase
and intensity, such as vibration noise on optical lattice mirror, because these noise frequencies
are mostly below $1 MHz$. In contrast, in order to achieve similar momentum width, the pulse
duration in Raman filter is usually about $2ms$, and noises with frequency higher than $500Hz$
will be received by the system. Dark states in VSCPT are much more fragile. Hence, our filter is
of great potential to achieve ultra-narrow momentum width because a lot of noises are shielded.

In conclusion, we apply a designed temporal matter wave Talbot-Lau
interferometer for realizing a momentum filter, which can keep the
atoms with lower momenta in their initial states and drive the
others to states with higher momenta. This kind of momentum filter
makes use of the momentum-state interference to discriminate between
momenta and hardly depends on the transition between the atomic
internal states. With such filter, we manage to prepare an atomic
cloud with  $0.056\hbar k$ momentum width within $300\mu s$. We
consider the interaction between atoms for condensate during the
filter process, and find that it affects the redistribution of
atomic gas, but hardly momentum purifying. We show that it is an
effective method for rapidly compressing the momentum width of an
atomic gas, and systematically discuss the specifications of the
filter. It may be of great potential for achieving ultra low
temperatures.

\section*{Acknowledgement}
We appreciate H. W. Xiong, G. V. Shlyapnikov, G. J. Dong, L. B. Fu,
B. Wu and H. zhai for stimulating discussions. We thank Thibault
Vogt for critical reading of our paper. This work is supported by
the National Fundamental Research Program of China under Grant No.
2011CB921501, the National Natural Science Foundation of China under
Grant No. 61027016, No.61078026 and No.10934010.

\end{document}